\documentclass[twocolumn, usenatbib, useAMS,usegraphicx]{mn2e}
\usepackage[english]{babel}
\usepackage{amssymb,amsmath}

\begin{document}

\newcommand{\gcc}{\mbox{${\rm g~cm^{-3}}$}}
\newcommand{\ks}{KS~1731--260}

\title[Neutron star cooling in KS~1731--260]{Neutron star cooling
after deep crustal heating in the X-ray transient KS~1731--260}
\author[P.~S.~Shternin et al.]
        {P.~S.~Shternin$^1$,  
        D.~G.~Yakovlev$^1$,   
    P.~Haensel$^2$,       
    and A.~Y.~Potekhin$^1$
    \\
        $^1$ Ioffe Physico-Technical Institute,
    Politechnicheskaya 26, 194021 Saint-Petersburg, Russia\\
        $^2$ N. Copernicus Astronomical Center, 
        Bartycka 18, PL-00-716 Warsaw, Poland }

\date{Accepted 2007 August 23. Received 2007 August 22; in original form 2007 August 01}

\pagerange{\pageref{firstpage} -- \pageref{lastpage}} \pubyear{2007}

\maketitle

\label{firstpage}

\begin{abstract}
We simulate the cooling of the neutron star in the X-ray transient
KS 1731--260 after the source returned to quiescence in 2001 from
a long ($\gtrsim 12.5$ yr) outburst state. We show that the cooling can be
explained assuming that the crust underwent deep heating during the
outburst stage. In our best theoretical scenario the neutron star
has no enhanced neutrino emission in the
core, and its crust is thin, superfluid, and has
the normal thermal conductivity. The
thermal afterburst crust-core relaxation in the star may be not over.
\end{abstract}


\section{Introduction}

\ks\ is a neutron star X-ray transient
whose observational history has been described
recently by \citet{Ca06}.
The source was discovered in the active state in August 1989
by the \textit{Kvant}
orbital observatory; subsequent analysis
showed that it had also been active in October 1988
\citep{Su90}.
It remained a bright X-ray source
showing type I X-ray bursts for about 12.5 years.
It is 
believed that
this activity was powered by accretion onto the neutron star
(through an accretion disk) from its low-mass companion
in a compact binary.
Adopting a distance to the source of $D=7$ kpc,
\citet{Ca06} report
a characteristic 2--10 keV luminosity of \ks\
in the active state of $\sim 10^{36}$
erg~s$^{-1}$. The source remained active
till the beginning of 2001
and then returned to quiescence.
The last detection in the active state was made
on January 21, 2001 with {\it RXTE},
but on February 7, 2001, {\it RXTE} 
already
failed to detect
\ks\ in the active state \citep{Wi01}.

The first detection of \ks\ in quiescence was made by \citet{Wi01}
with {\it Chandra} on March 27, 2001. For $D=$7 kpc,
the 0.5--10 keV luminosity was
$\sim 10^{33}$ erg~s$^{-1}$, three orders of magnitude lower
than in the active state. The radiation spectrum
contains a component that can be
interpreted as the thermal emission from the neutron star
surface.
Since then the source has been
observed several times with {\it Chandra} and {\it XMM-Newton}
as summarized by \citet{Ca06}. Its X-ray light curve
faded over time scale $\sim 2$ years showing
a trend to flattening (with the residual luminosity
of $\sim 2 \times 10^{32}$ erg~s$^{-1}$).

According to observations, the accretion in quiescent states
of X-ray transients is stopped or strongly suppressed.
The nature of quiescent X-ray emission is a subject
of debates (see, \citealt{Ca06} and references therein,
for a list of possible
hypotheses). Here, we focus on the hypothesis of deep
crustal heating of neutron stars proposed by \citet*{BBR98}.
It states that, when a neutron star
accretes,
its 
crust is heated by nuclear transformations
(mainly by beta captures and pycnonuclear reactions)
in the accreted matter sinking within the crust under
the weight of newly accreted material.
The star remains sufficiently warm after an accretion
episode,
producing quiescent surface emission.
The sequence
of nuclear transformations and associated energy generation
rates were calculated by \citet{HZ90} assuming that the accreted matter
burns to $^{56}$Fe in the neutron star surface layers
so that, initially, before
sinking within the deep crust, the matter is composed of $^{56}$Fe.
Later \cite{schatz01} calculated explosive
nucleosynthesis in the neutron star surface layers
and showed that the explosive burning can proceed to much heavier
elements. Accordingly, \citet{HZ03} proposed
new deep crustal heating  scenarios (starting
with heavier elements, particularly, with $^{106}$Pd).
In all the cases \citet{HZ90,HZ03} obtained similar deep crustal
energy release, $\sim 1-1.5$ MeV per one accreted nucleon,
sufficient to power quiescent thermal emission in X-ray transients.
Recently \citet{Gu07} have reconsidered the heating 
starting
with multicomponent matter (ashes of explosive burning
in the surface layers). They have shown that the heating
of the deep outer crust
can be higher because beta captures
can produce daughter nuclei in excited states;
their deexcitation can 
generate
extra heat.

The onset of the quiescent state of \ks\ was recognized
as an outstanding phenomenon from the very beginning.
The majority
of other
X-ray transients undergo short accretion
episodes (days to months) in which the deep crustal heating
cannot 
break the crust-core thermal coupling and
make the crust much hotter than the 
stellar
core.
However, it is
possible in \ks\ because of the long accretion stage
\citep{Ru02}.
Therefore, observations of its
quiescent thermal emission can help to understand how the
crustal heat spreads over the entire star,
that is useful for exploring
the neutron star structure.

The first modelling of
the \ks\ cooling was made by \citet{Ru02} soon
after the quiescence onset.
The authors based on previous simulations
by \citet{UR01}
of the crust-core relaxation
in a neutron star with a heated crust.
\cite{Ru02} proposed several cooling scenarios
based on the deep crustal heating model of
\citet{HZ90} and
different crust and core microphysics.
They predicted that the neutron star can reach the crust-core
relaxation
and associated flattening of the quiescent soft X-ray light curve
in 1--30 years.
\citet{Ca06} have compared the new
observations of \ks\ with the predictions of \citet{Ru02} 
and conclude that the star
should have high thermal conductivity in the crust and
enhanced neutrino emission in the core.

Here we present new cooling calculations
and discuss their consistency with the observations of \ks.

\section{Cooling model and physics input}
\label{S:cool}

\begin{figure}
\includegraphics[width=8cm, bb=20 165 575 595]{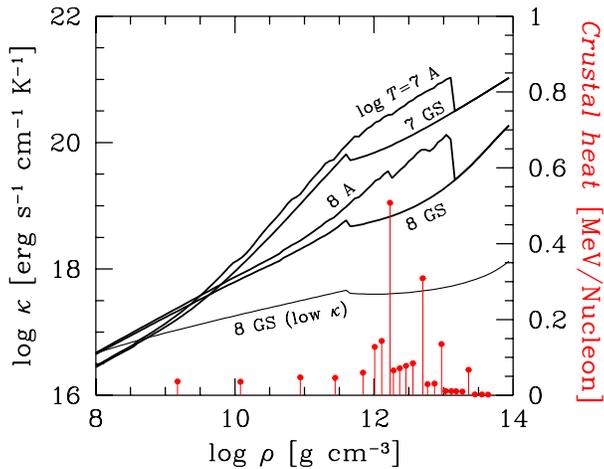}
\caption{Density dependence of the electron
thermal conductivity $\kappa$ (left vertical scale) in the neutron star crust
with accreted (A) or ground-state (GS) matter
at two temperatures ($\log T$[K]=7 and 8,
numbers next to curves). The thin lower curve is
for the model of low $\kappa$ while other curves are for
normal $\kappa$.
Vertical bars
show positions and power (right vertical scale) of the heat
sources. Initial layer is assumed to consist
 of $^{56}{\rm Fe}$, as in \citet{HZ90}, but
 neutrino losses 
in 
electron captures 
are
suppressed, following \citet{Gu07}.
}
\label{F:thermcond}
\end{figure}

Our cooling simulations are similar to those of
\citet{UR01} and \citet{Ru02}.
We assume that the neutron star crust
underwent deep crustal heating during the long
accretion stage.
We employ the model of deep crustal heating of \citet{HZ90}, but
modify the energy release due to sequences of
pairs of beta-captures in
the crust.
Specifically, we assume that daughter nuclei after a
primary beta capture are produced in excited states
and deexcite before a secondary beta capture, 
heating
thus the matter (instead of wasting extra energy into
neutrino emission). In this way the distribution of
heating sources remains the same as
in \citet{HZ90} but the sources in the outer crust become stronger,
resembling those obtained by \citet{Gu07}.
The source positions and strengths,
calculated by \citet{HZ07}, are
shown in Fig. \ref{F:thermcond}.
The overall energy release is
$1.9$~MeV per accreted nucleon.

To simulate the neutron star cooling we use our general relativistic
cooling code (\citealt*{Gn01}). It solves the thermal diffusion
problem within the star (at densities $\rho> \rho_\mathrm{b}$)
and uses a predetermined quasi-stationary relation $T_\mathrm{s}-T_\mathrm{b}$
\citep*{Palex97} between
the effective surface temperature $T_\mathrm{s}$ and
the temperature $T_\mathrm{b}$ at 
the base
($\rho=\rho_\mathrm{b}$) of a thin heat-blanketing envelope
($\rho \leq \rho_\mathrm{b}$). Now we 
shift 
$\rho_\mathrm{b}$
from previously used values $\sim 10^{10}-10^{11}$ g~cm$^{-3}$
to $\rho_\mathrm{b}=10^8$ g~cm$^{-3}$. This allows us to
put
all heat sources into the region of $\rho > \rho_\mathrm{b}$
and to reduce the time of heat propagation through the blanketing
layer from $\sim 1$ yr to $\sim 1$ d (enabling the code to
trace short-term 
-- 1 day --
surface temperature variations).

To explore the sensitivity of calculations to the crust physics,
we employ two models of the neutron star crust, composed of
ground-state (GS) or accreted (A) matter. The ground-state crust
(e.g., \citealt{NSB}) has been used in our previous simulations.
The model of accreted crust \citep{HZ90} is consistent
with the adopted model of deep crustal heating.
The accreted crust is composed of lighter nuclei with
lower atomic numbers.
Deep in the inner crust, at $\rho \gtrsim 10^{13}$
g~cm$^{-3}$,
composition is similar
to the ground-state one,
with 
$\gtrsim$80\%  of nucleons constituting
a neutron gas 
\citep{HZ90}.

We employ the electron thermal conductivity in the crust, $\kappa$,
limited by
electron-ion 
\citep{Gn01}
and 
electron-electron 
\citep{SY06}
scattering. It will be called {\it normal}.
We will also use the model electron thermal conductivity
proposed by \citet{brown00}. It corresponds to
an amorphous crust \citep[e.g.,][]{jones04mn}
 and will be called {\it low}.
Actually, it gives the lowest limit on $\kappa$ in the crust.
Several model thermal conductivities as functions of density
in the crust for
two values of temperature ($T=10^8$ and $10^7$~K)
are plotted in Fig.~\ref{F:thermcond}.

In the inner crust, we take into account the effects
of neutron superfluidity on the
heat capacity of free neutrons (e.g., \citealt*{Yak99}).
A representative set of models for superfluid neutron gaps
in
the inner crust, which determine
superfluid critical temperature profiles $T_\mathrm{c}(\rho)$,
is collected by \citet{LS01}.
The collection includes a well defined gap provided by
the pure BCS theory of singlet-state neutron pairing
(with a maximum of $T_\mathrm{c} \sim 2 \times 10^{10}$~K
within the crust) and a number of gaps calculated
using various neutron polarization models (with
the maxima of $T_\mathrm{c}$ approximately three times lower).
BCS superfluidity very strongly suppresses the neutron heat
capacity in the inner crust; this superfluidity will be called {\it strong}.
The effects of other
superfluid models are weaker and more or less similar. For illustration
of the latter effects,
we will use the model proposed by \citet*{Wa93};
such superfluidity will be called {\it moderate}.
We calculate the neutrino emission in the crust and in the core 
according to \citet{Yak-ea01}.
 In our cooling models
the neutron star 
stays
not too hot, so that crustal neutrino emission
(including that due to Cooper pairing of neutrons)
is insignificant.

In the neutron star core, we use an equation of state of dense
matter (containing nucleons, electrons, and muons) constructed
by \citet*{APR98} (their model Argonne V18+$\delta v$+UIX$^*$).
Specifically, we adopt its convenient
parametrization proposed by \citet{HH99} and described as APR~I
by \citet{Gus05}. 
In this case,
the 
maximum 
gravitational
mass of stable neutron
stars is $M_\mathrm{max}=1.923 M_\odot$ and
the direct Urca process of powerful neutrino emission opens at
$M>1.828 M_\odot$. We will mainly use two neutron star
models, with 
masses $M=1.6$ and $1.4 \,M_\odot$, where
direct Urca process is forbidden; both stars demonstrate slow
neutrino cooling via the modified Urca process.
The $1.4\,M_\odot$ star has the central density
$\rho_\mathrm{c}=9.4 \times 10^{14}$ \gcc, the
circumferential radius $R=12.14$ km, and the crust
thickness $\Delta R=R-R_\mathrm{core}=1.16$ km
(where $R_\mathrm{core}$ is the core radius corresponding to
$\rho=1.5\times 10^{14}$ \gcc). The $1.6\,M_\odot$ star
is more compact, with thinner crust,
and has
$\rho_\mathrm{c}=1.16 \times 10^{15}$~\gcc, 
$R=11.88$~km, and $\Delta R=890$~m.

The thermal conductivity of the neutron star core
is described following
\citet*{Ba01} and \citet{SY07}. For simplicity, the effects of
nucleon superfluidity in the core are neglected.

\vspace*{-1ex}
\section{Results and Discussion}
\label{S:res}

\begin{figure*}
\includegraphics[width=16cm, bb=20 10 580 217]{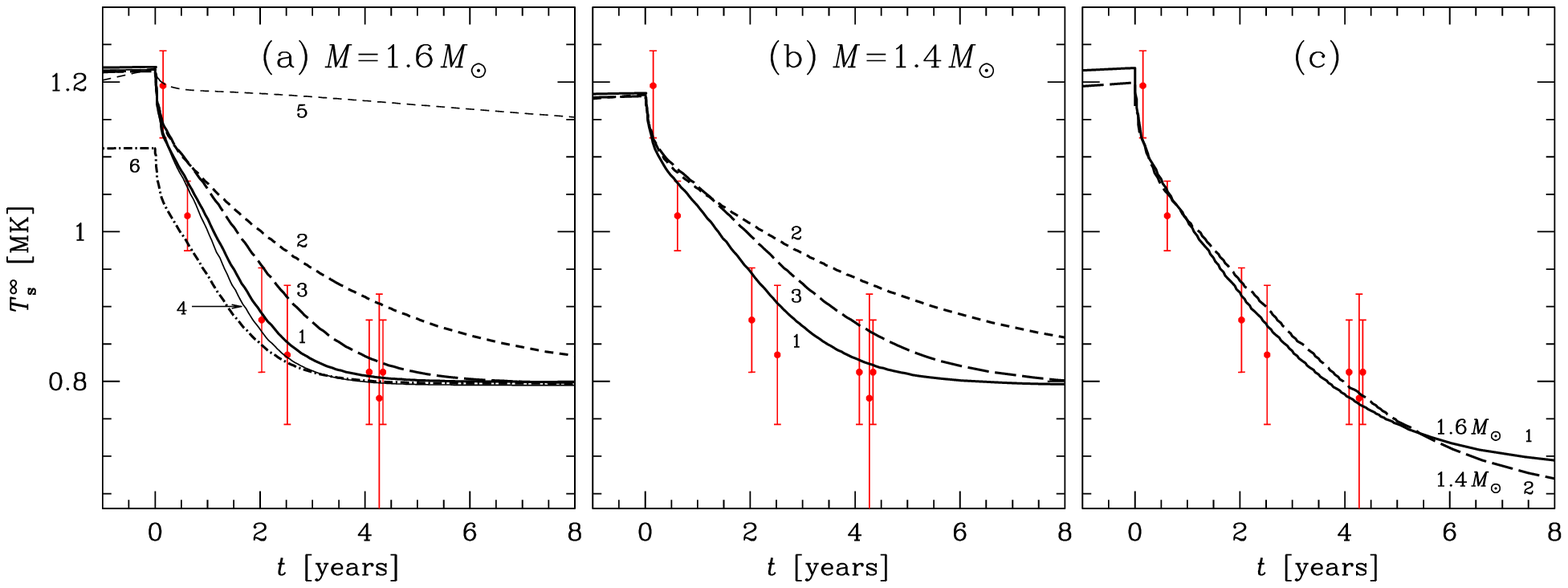}
\caption{Theoretical cooling curves for
(a) $M=1.6\,M_\odot$ and (b) $1.4\,M_\odot$
neutron stars, and (c) for stars with both $M$
compared with observations. The curves are explained
in Table \ref{tab:curves} and in text. 
}
\label{fig:cool}
\end{figure*}

We have calculated (Fig.\ \ref{fig:cool}) a number of cooling curves
which give
the effective surface temperatures $T_\mathrm{s}^\infty$,
as detected by a distant observer, versus time $t$; $t=0$
refers to
February 1, 2001, the date near which \ks\ turned in
quiescence. We compare the curves
with seven observational points presented by \citet{Ca06};
the values of $T_\mathrm{s}^\infty$
were inferred from 
the
observed X-ray spectra
(employing non-magnetic neutron star
hydrogen atmosphere models from the Xspec database
and assuming $D=7$ kpc, $R=10$ km,
and $M=1.4\,M_\odot$).
We doubled 
the reported
$1\sigma$ observational
error bars to 
enlarge
statistical significance,
that would make
our analysis more realistic. 

To start
any
 cooling calculation, we have taken a neutron star model
with the thermally relaxed interior and some
initial surface temperature $T_\mathrm{s0}^\infty$.
Then we switch on deep crustal heating produced by
a constant mass accretion rate $\dot{M}$ over 12.5 years.
In that period a certain amount of heat $E_\mathrm{tot}$
is deposited into the crust. 
The crust is heated
and
its thermal balance with the thermally inertial
core
is violated.
Then we switch off accretion
(deep crustal heating) and the crust cools down regaining
thermal equilibrium with the core.
Some (typically small)
 part of $E_\mathrm{tot}$
diffuses to the surface and radiates away via thermal
surface emission. The rest is carried by thermal
conduction to the core. The core temperature stays
almost unchanged because of the high core thermal
conductivity and heat capacity.
The crust-core
thermal relaxation takes 1--100 years, depending on
the neutron star model. After this relaxation is over,
the surface temperature nearly reaches its initial value
$T_\mathrm{s0}^\infty$.
The star 
cools down further 
with isothermal interior
over typical cooling time scales
1--10 kyr
until the next accretion episode. The extra heat
deposited to the core is mainly emitted
over those long cooling time scales
via core neutrino emission.

Our cooling curves in Fig.\ \ref{fig:cool} are calculated
for different neutron star masses, microphysics in the crust,
mass accretion rates (and $E_\mathrm{tot}$), and $T_\mathrm{s0}^\infty$
(as shown in the figure and Table \ref{tab:curves}).
Our aim is to explain the observed temporal evolution
$T_\mathrm{s}^\infty(t)$ of \ks\ in the quiescent
state.
A successful explanation should also be consistent with
the observational constraint on the mass accretion rate,
$\dot{M}\lesssim 5\times 10^{-9}M_\odot$~yr$^{-1}$ (for
$D=8$~kpc, see Table 3 in \citealt*{Yak03}), which translates
into $E_\mathrm{tot}\lesssim 2.4\times 10^{44}$~erg for
the adopted deep heating model. Table \ref{tab:curves}
shows that all presented cooling models roughly satisfy this
requirement.

Fig.\ \ref{fig:cool}a refers to the $1.6\,M_\odot$ neutron star model,
while Fig.\ \ref{fig:cool}b is for the $1.4\,M_\odot$ star.
All curves in Figs.\ \ref{fig:cool}a and b are calculated assuming
the initial surface temperature 
to be
$T_\mathrm{s0}^\infty=0.8$~MK
(so that the internal temperature is $\sim 8 \times 10^7$~K,
as in a cooling isolated neutron star which is $\sim 10^5$ years old).
This is a typical surface temperature of the neutron star
provided by the last three observational points.
Thus, in
Figs.\ \ref{fig:cool}a and b we (following \citealt{Ca06})
tacitly assume that the crust-core equilibrium is
re-established
in two years after the quiescence onset.
In all curves in Figs.\ \ref{fig:cool}a and b, but in curve 6,
$E_\mathrm{tot}$
 has been chosen in such a way for
the surface temperature at the first quiescent observation 
to be 
consistent with data.

Curve 1 in Fig.\ \ref{fig:cool}a seems to be the best.
It corresponds to the accreted crust with the normal
thermal conductivity and moderate neutron superfluidity.
It naturally explains the thermal relaxation of \ks\ with
the standard physics input.
The maximum
internal temperature raise to $T \sim 4 \times 10^8$~K
takes place at $t=0$
near the boundary between the outer and the inner crust.
The core-crust relaxation takes $\sim 2$ years.
The star would need
$\sim10^3$ years to emit all the heat pumped into the core 
during the outburst and
reach the same
thermal state as before the outburst. This is in good agreement
with the estimate of
\citet{Ru02}
 for a similar cooling model.

Using the same physics as for curve 1 but the ground-state
crust (with lower conductivity) we obtain slower relaxation
(curve 3). It is acceptable but less consistent with
the observations; it requires lower $E_\mathrm{tot}$ because
it is easier to heat 
the crust with
smaller thermal conductivity.
Taking the latter cooling model 3 and
neglecting superfluidity in the inner crust we obtain curve 2.
A non-superfluid crust has larger (neutron) heat capacity
which
noticeably delays the thermal
relaxation making it much less consistent with the data.
Returning to our best model 1 but assuming strong superfluidity
in the crust,
 we stronger suppress the heat capacity of neutrons
and obtain curve 4; it shows faster and quite acceptable relaxation.
The effects of strong and moderate superfluidity are actually very
close, although the presence of superfluidity greatly improves 
the agreement with the data.
Now if we return to model 1 but assume low thermal conductivity,
we come to much longer
crust-core relaxation (over several hundred years, curve 5). It is inconsistent
with 
the
observations, in agreement with the
conclusion of \citet{Ca06}. Finally, if we take the best model 1
but assume the same (lower) mass accretion rate as in model 2,
we get curve 6. Therefore, the
latter
 mass accretion rate,
being used for the microphysics
 of model 1,
is insufficient
to explain high values of $T_\mathrm{s}^\infty$ in the
beginning of the quiescent state.

Curves 1--3 in Fig.\ \ref{fig:cool}b are analogous to
curves 1--3 in Fig.\ \ref{fig:cool}a, but are calculated
for a less massive star, with thicker crust.
The
thicker
crust produces longer thermal relaxation, less
consistent with the observations.

\begin{table}
\caption[]{Cooling curves in Fig.\ 2}
\label{tab:curves}
\begin{center}
\begin{tabular}{  c  c  c  c  c  c}
\hline
\hline
 Curve & $T_\mathrm{s0}^\infty$ & Crust & Conduction & 
Superfluid & $E_\mathrm{tot}$ \\
       &  MK  & model & in crust & in crust & $10^{44}$ erg \\
\hline
\hline
 1a & 0.8 & A  & normal & moderate & 2.6  \\
 2a & 0.8 & GS & normal & none   & 1.9  \\
 3a & 0.8 & GS & normal & moderate & 1.8  \\
 4a & 0.8 & A  & normal & strong & 2.6  \\
 5a & 0.8 & A  & low   & moderate & 0.6 \\
 6a & 0.8 & A  & normal & moderate & 1.9  \\
\hline
 1b & 0.8 & A  & normal & moderate & 2.3  \\
 2b & 0.8 & GS & normal & none   & 1.7  \\
 3b & 0.8 & GS & normal & moderate & 1.5  \\
\hline
 1c & 0.67 & GS  & normal & none & 2.4  \\
 2c & 0.63 & GS  & normal & none & 2.4  \\
\hline
\end{tabular}
\end{center}
\small{
}
\end{table}

We have also performed many other cooling calculations
varying physics input.
In particular, we have varied the distribution of heat sources within
the crust. We have obtained that
 it is much easier to
explain the observations by placing the sources into the
outer crust. These models naturally give
short thermal relaxation and efficient heating of the surface.
In contrast, were all sources located in the deep inner
crust, the star would show longer thermal relaxation
and one would need too much energy to heat the surface because
the heat would be pumped into the core.
In
connection to this,
the improved model of deep crustal heating
used here,
where the heat release in
the crust
is
enhanced compared to the original model
of \citet{HZ90} (due to switching-off
neutrino losses associated with electron captures,
\citealt{Gu07}),
is more favorable for
explaining the observations.

In addition, we have artificially varied the thermal conductivity in
different places of the crust and found high sensitivity
of the cooling curves to these variations.
The conductivity strongly affects both, the thermal
relaxation time 
and the efficiency of surface heating.
Taking the conductivity a few times lower
than the normal conductivity of accreted or ground-state
crust produces too long crust-core relaxation which
disagrees with the data.

Furthermore, we have taken different neutron star models
(different equations of state in the core, and different masses).
In particular, we have used the models of massive neutron
stars whose core neutrino emission is strongly enhanced
by the nucleonic direct Urca process (e.g., $1.9\,M_\odot$ model
for the equation of state employed in Fig.\ \ref{fig:cool}).
We have found that we need unrealistically intense crustal heating
(too high $E_\mathrm{tot}$) to explain the high observed
values of $T_\mathrm{s}^\infty(t)$ in the beginning
of quiescence. Moreover, such a star
has too short global cooling time scale (years to
decades),
comparable to the crust-core relaxation time. 
The
crust-core relaxation 
becomes
coupled to
the global thermal relaxation; the cooling curves
do not show the observed flattening at $t \gtrsim 2$ years.
Hence, we cannot reconcile theory with observations if
the neutrino emission of the star is enhanced by the direct
Urca process. Nevertheless, we think that it 
may
be
possible to explain the observations if the neutrino emission
is enhanced by a less efficient mechanism (e.g., by pion or
kaon condensation in the stellar 
core)
or if the direct Urca process operates but is strongly
suppressed by nucleon superfluidity
(e.g., \citealt{YP04}, \citealt*{Pa06}).

Finally, we remark that
the thermal crust-core
relaxation in \ks\ may be still not over.
This 
is illustrated in Fig.\ \ref{fig:cool}c,
where we present two new cooling curves for our $1.6\,M_\odot$
and $1.4\,M_\odot$ neutron star models.
They are calculated without imposing the constraint
that 
$T_\mathrm{s0}^\infty=0.8$~MK.
We have intentionally taken
the physics input (ground-state, non-superfluid crust
with normal conductivity) which gives too long
thermal relaxation to
explain the data for the scenarios in Figs.\ \ref{fig:cool}a and b.
Now we
take
lower $T_\mathrm{s0}^\infty$ and
reach consistency with the current observations
(and
get
a rather
low crustal heat release $E_\mathrm{tot}$).
We see
that 
the crust-core
relaxation in \ks\ can 
really
last longer than 2 years, and this 
possibility
widens
the class of cooling models consistent with the data.
It
will hopefully be checked in future
observations of \ks.

\section{Conclusions}

We have simulated the cooling of the neutron star
in the quiescent state of \ks\
employing the model of deep crustal heating of the star
in the outburst state.
We have used the model of deep crustal heating
\citep{HZ90} updated
by switching-off neutrino losses
in the crust
\citep{Gu07}. Our main
conclusions are:
\begin{enumerate}

\item One can explain current observations
of \ks\ 
using a model of deep crustal heating and
a standard microphysics of the neutron star.

\item If the crust-core thermal relaxation in the neutron star
is reached in $\sim 2$ years, the most successful cooling
model implies the model of accreted crust with normal
thermal conductivity and neutron superfluidity;
the neutron star should be sufficiently massive
(to have a thinner crust),
 but the neutrino emission
in its core 
cannot be too high
(e.g., 
it can be
provided by the
modified Urca 
process).
All these factors shorten
the crust-core thermal relaxation.

\item The model of low thermal conductivity
(amorphous crust)
gives too long crust-core relaxation, inconsistent
with the data.

\item The enhanced neutrino cooling via the direct
Urca process in the neutron star core gives too
fast cooling of the entire star and requires too intense crustal
heating, inconsistent with the data.

\item The crust-core thermal relaxation can be not reached
yet. If so, the data can be explained by
a
 wider
class of 
neutron star
models.

\end{enumerate}

We stress that the thermal crust-core relaxation of
the neutron star in \ks\ is much more sensitive
to the physics of the crust than the core.
We employed the models of non-superfluid core just for
simplicity. Core superfluidity can change the
core heat capacity and neutrino luminosity,
 but
the
principal
conclusions will be the same.
Our calculations are not entirely self-consistent.
For instance,
 the surface temperature was inferred
from observations \citep{Ca06},
 assuming neutron star
masses and radii different from those used in our
cooling models. 
This
inconsistency cannot affect
our main conclusions,
 but
 it would be desirable to infer
$T_\mathrm{s}^\infty$ for our neutron star models.
The thermal relaxation in the quiescent state has been
observed also \citep{Ca06} for another neutron star X-ray transient,
MXB 1659--29. We hope to analyse these data in the
next publication.

\section{acknowledgments}
We are grateful
to A.~I.\ Chugunov, O.~Y.\ Gnedin, K.~P.\ Levenfish
and Yu.~A. Shibanov for useful discussions,
and to the referee, Ulrich Geppert, for valuable remarks.
The work was partly supported by the
Russian Foundation for Basic
Research (grants 05-02-16245, 05-02-22003),
by the Federal Agency
for Science and Innovations
(grant NSh 9879.2006.2),
and by the Polish Ministry of Science and
Higher Education (grant N20300632/0450).
One of the authors (P.S.) acknowledges
support of the Dynasty Foundation
and perfect conditions of the Nicolaus Copernicus Astronomical
Center in Warsaw, where this work was partly performed.
P.H.\ and A.P.\ thank the Institute for Nuclear Theory 
at the University of Washington for 
hospitality and the
U.S.\ Department of Energy for partial support
during the completion of this work.

\label{lastpage}

\end{document}